\documentclass[preprint2]{aastex}
\begin{document}


\shorttitle{Gamma-ray spectra from dense gas clouds}
\shortauthors{Ohishi et al.}

\title{Gamma-ray spectra due to cosmic-ray
interactions with dense gas clouds}

\author{Michiko Ohishi, Masaki Mori}
\affil{Institute for Cosmic Ray Research, University of Tokyo,
Kashiwa, Chiba 277-8582, Japan}
\author{\&}
\author{Mark Walker}
\affil{School of Physics, University of Sydney, NSW2006, Australia}
\affil{Australia Telescope National Facility, CSIRO, Australia}

\begin{abstract}
Gamma-ray spectra from cosmic-ray proton and electron interactions
with dense gas clouds have been calculated using a Monte Carlo 
event simulation code, GEANT4. Such clouds are postulated as
a possible form of baryonic dark matter in the Universe.
The simulation fully tracks the cascade and transport processes
which are important in a dense medium, and the resulting
gamma-ray spectra are computed as a function of cloud column-density.
These calculations are used for predicting the Galactic
diffuse gamma-ray spectrum which may be contributed by baryonic
dark matter; the results are compared with data from the
EGRET  instrument, and used to constrain the
fraction of Galactic dark matter which may be in the form of
dense gas clouds. In agreement with previous authors, we find
useful constraints on the fraction of Galactic dark matter which
may be in the form of low column-density clouds
($\Sigma\la 10\,{\rm g\,cm^{-2}}$). However, this fraction rises steeply
in the region $\Sigma\sim10^2\,{\rm g\,cm^{-2}}$, and for
$\Sigma\ga200\,{\rm g\,cm^{-2}}$ we find that baryonic dark matter
models are virtually unconstrained by the existing gamma-ray data. 
\end{abstract}


\section{Introduction}
The nature of dark matter remains one of the outstanding
questions of modern astrophysics. The success of the
cold dark matter cosmological model (albeit with ``dark energy''
now required: $\Lambda$CDM) argues strongly for a major component
of the dark matter being in the form of an elementary particle.
However, the inventory of baryons which we can observe
locally falls far short of the total inferred from observations
of the cosmic microwave background fluctuations
\citep{Fuk04}, leaving
open the possibility that there may be a significant baryonic
component of dark matter. Furthermore, although $\Lambda$CDM
is very successful in describing the growth of structure in the
universe on large scales, we still lack a direct detection of any
of the candidate dark matter particles. Lacking this decisive
piece of observational evidence, some authors have proposed
models which include a large component of baryonic dark matter.
In particular there have been many papers dealing with the
possibility that cold, self-gravitating molecular clouds
constitute a major component of the dark matter
\citep{Pfe94,DeP95,Hen95,Ger96, Com97, Wal99, Sci00a, Sci00b}.
A variety of different forms, including isolated, clustered,
and fractal, have
been considered for the clouds, but all proposals involve dense
gas of high column-density, in contrast to the diffuse gas in the
interstellar medium which is easily detected in emission and/or
absorption. 

One of the fundamental predictions of a model featuring dense
gas clouds is the
gamma-ray emission resulting from cosmic-ray interactions
within the clouds \citep{DeP95, Kal99, War99, Sci00a}.
Because of the potentially large total
mass of gas involved, this process may yield a diffuse flux
in the Galactic plane comparable to the flux from known
sources for photon energies around 1 GeV \citep{Sci00a}.
Considering the high quality data on diffuse emission
acquired by the EGRET detector aboard the Compton Gamma
Ray Observatory \citep{Hun97},  it is worth considering this
source of gamma-ray emission in detail as it is possible to use
these data to constrain the dark matter models (see \citet{Sal96};
 \citet{Gil94}).  Most previous investigations of this problem
have neglected the self-shielding and cascade phenomena
which can be important at high column densities \citep{Kal99, Sci00a},
and have employed emissivities appropriate to the low-density
limit. These effects alter the emergent gamma-ray spectrum,
and we note that this could be relevant to the
observed excess Galactic flux above 1 GeV \citep{Hun97}.
We have noted elsewhere \citep{Wal03} that massive
($M\ga10^6\;{\rm M_\odot}$) aggregates of dense
gas clouds could potentially account for many of the
unidentified discrete sources detected by EGRET \citep{Har99}.

Here we present detailed calculations of the gamma-ray
spectra arising from cosmic-ray interactions with dense
gas clouds. We have used a Monte
Carlo simulation code, GEANT4, developed for simulating
interaction events in detectors used in high-energy particle physics.
Not surprisingly, we find that the predicted spectra differ
substantially between high and low column-density clouds, and
we discuss the interpretation of our results in the context
of the observed Galactic gamma-ray emission. Our
calculations are undertaken for cold, dense molecular gas
in clouds of radius $R\sim 10^{13}$cm, similar to those
proposed by \citet{Wal98} to explain the
extreme scattering events \citep{Fie94} during which
compact extragalactic radio sources are magnified
and demagnified as a plasma ``lens'' moves across the line 
of sight (see \citet{McK01} for a criticism of this model).  
However, the results of our calculations depend primarily on
the column-density of the individual clouds,
$\Sigma$, under consideration, and their
fractional contribution to the Galaxy's dark matter halo, and
our results can be taken as representative of other models
which are characterised by similar values of these quantities.

\section{Results}

\subsection{Gamma-ray production in dense gas}
Previous calculations of gamma-ray spectra from
cosmic-ray irradiation assumed single interactions
of protons with the interstellar medium (\citet{Ste81, Der86, Ber93}
and references therein). In order to investigate cosmic-ray
interactions with dense gas, where cascade processes and
particle transport are important, we have used a Monte Carlo
code, GEANT4,\footnote{Available at
http://wwwinfo.cern.ch/asd/geant4/geant4.html} 
to derive gamma-ray production spectra.
This code is a general purpose Monte Carlo code of particle
interactions and is widely used for simulation of high-energy
particle detectors in accelerator experiments.
Cross-sections and interactions of various hadronic
processes, i.e., fission, capture, and elastic 
scattering, as well as inelastic final state production,
are parametrized and extrapolated in high and low particle
energy limits, respectively.
The $\pi^0$ production in this code, which is important because
of the $\pi^0\rightarrow2\gamma$ decay that dominates the
emissivity of the gas at high energies, has been tested
against accelerator data \citep{Dan99}.

Initially we experienced one slight difficulty in applying
GEANT4 to our physical circumstance:
the low-energy hadron interaction code, called GHEISHA,
did not conserve energy very accurately (GEANT4 bug reports
No.\ 171 and 389). A ``patch'' was available for GHEISHA
(GEANT ver.\  4.4.1), but this patch appeared to introduce
further problems of its own in the energy deposition
distribution (GEANT4 bug report No.\ 415). These difficulties
have been overcome by the GEANT team, and we are not aware
of any such problems in the latest release (GEANT ver.\ 4.5.1).

\begin{figure} 
\plotone{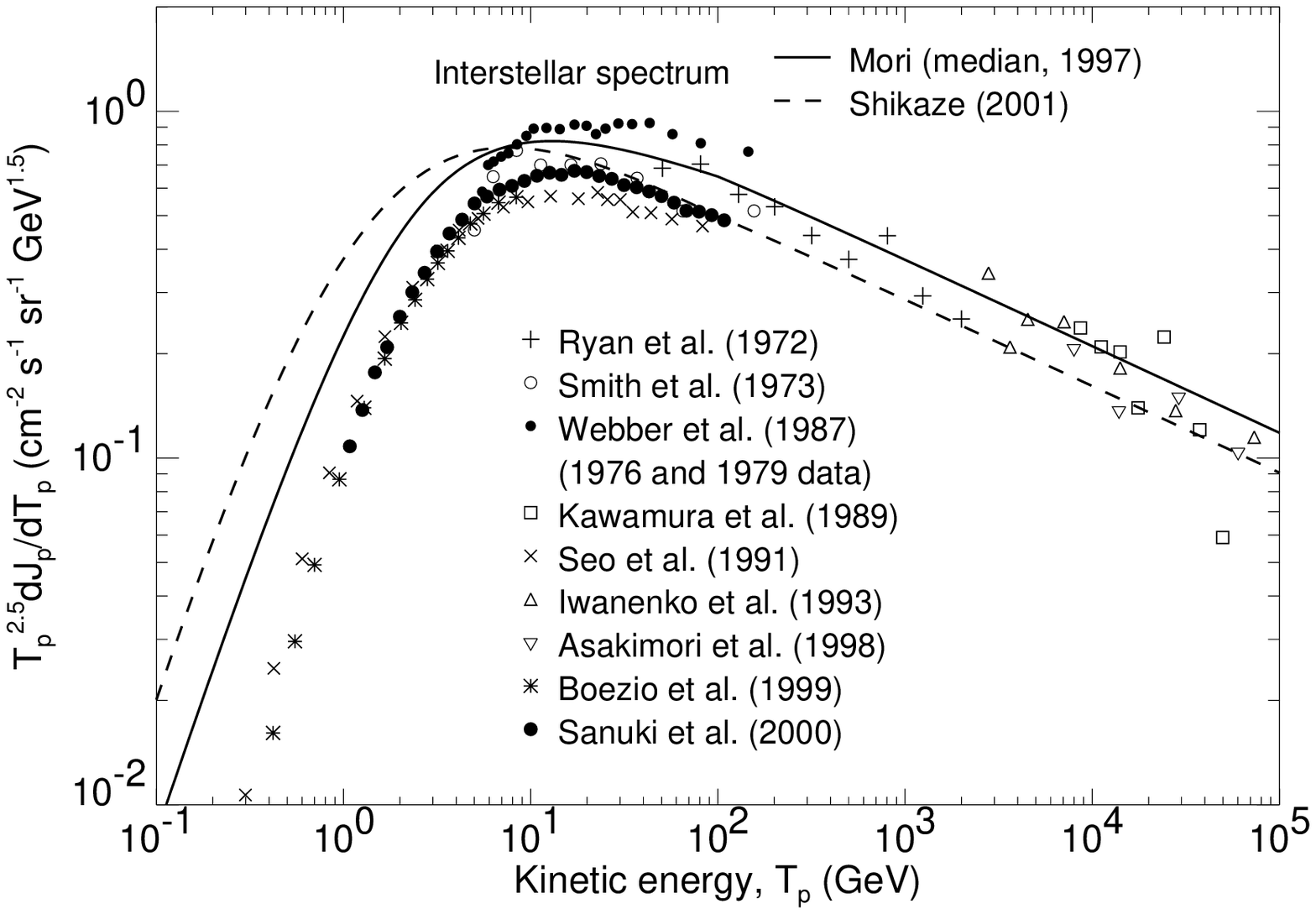}
\caption{Adopted cosmic-ray proton spectrum in the Galaxy.
The model curves are taken from \citet{Mor97} and \citet{Shi01}.
Also plotted are direct observations of the local spectrum
\citep{Rya72,Smi73,Web87,Kaw89,Seo91,Iva93,Asa98,Boe99,San00},
 but note that these are strongly affected by solar modulation.
\label{fig:f1} }
\end{figure}

\begin{figure} 
\plotone{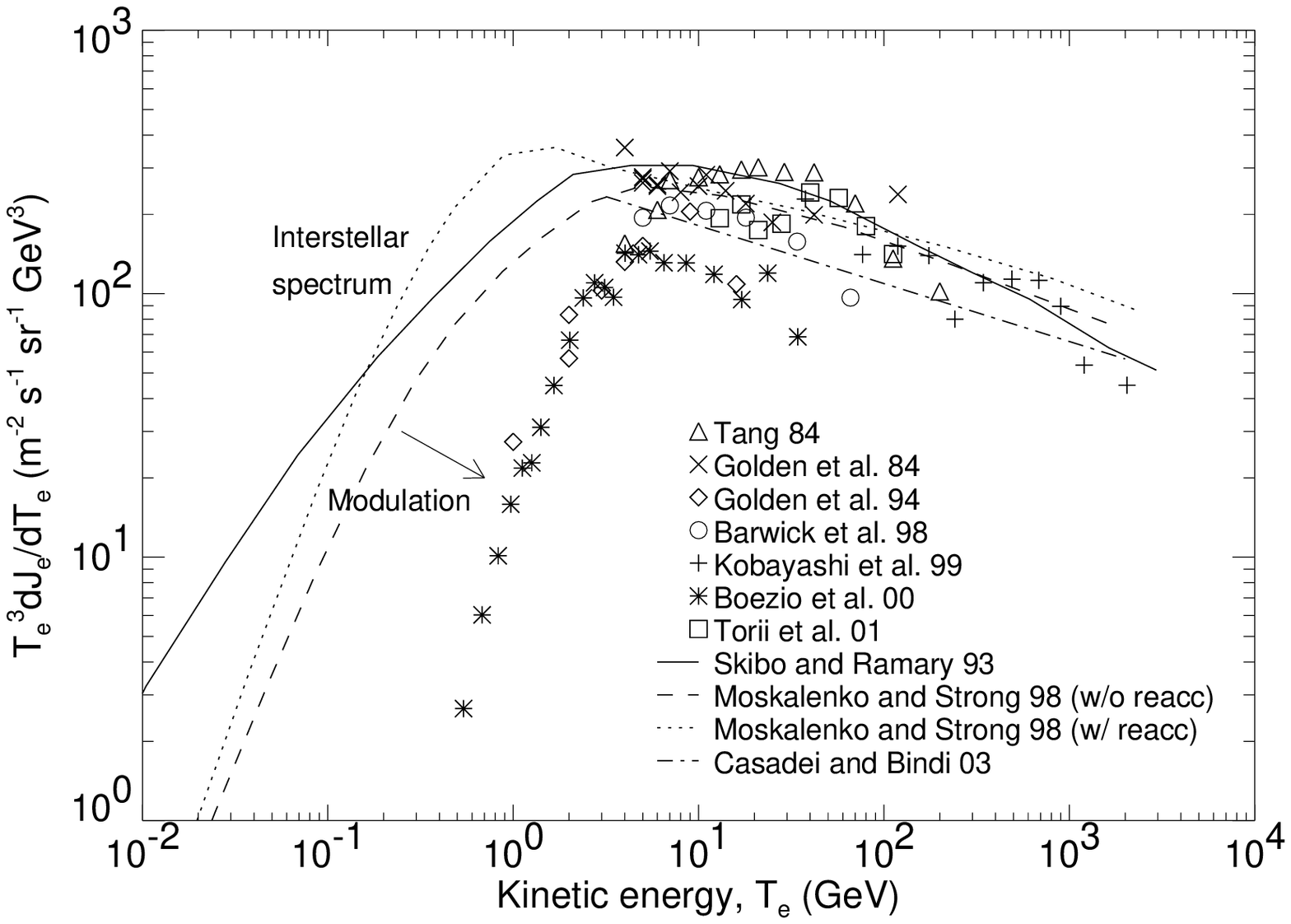}
\caption{Adopted cosmic-ray electron spectrum in the Galaxy.
The model curves are taken from \citet{Ski93}, \citet{Mos98} and
\citet{Cas03}.
(The two lines for \citet{Mos98} show models with and without 
reacceleration of the cosmic rays.)
Also plotted are direct observations of the local electron
spectrum \citep{Tan84,Gol84,Gol94,Bar98,Kob99,Boe00,Tor01},
 but note that these are affected by solar modulation.
\label{fig:f2} }
\end{figure}

Our calculations assume a spherical cloud of molecular
hydrogen of uniform density and temperature (10~K).
The radius of the sphere was assumed to be 
$R=1.5\times 10^{13}\;{\rm cm}\simeq1$~AU.
Protons and electrons are injected randomly at a
surface point of the cloud and particles subsequently emanating
from this surface are counted as products.
The adopted spectra of cosmic-ray protons and electrons
were taken from \citet{Mor97} (here we use the ``median'' 
flux; note that the units on his equation (3) should
read cm$^{-2}$s$^{-1}$sr$^{-1}$GeV$^{-1}$), and
\citet{Ski93}, respectively. In Figs.~\ref{fig:f1} and
\ref{fig:f2} we plot these spectra together with
observational data. (Note that the adopted spectra are
those for cosmic rays in the Galaxy, whereas the
measured points are subject to the modulating influence
of the magnetic field of the solar wind).
The simulated range of kinetic energy of cosmic rays
is from 10 MeV to 10 TeV. We divided this energy
range into four and superposed the resulting spectra
with appropriate weight factors
in order to increase the simulation statistics at
higher energies, considering the rapidly falling
spectrum of cosmic rays.
The density of molecular hydrogen, $\varrho$, was varied from
$5\times 10^{-16}$ to $5\times 10^{-9}$ g~cm$^{-3}$ 
in factors of 10. This corresponds to the column density,
$\Sigma=2\varrho R\langle\cos\theta\rangle$, of
$10^{-2}, 10^{-1}, \ldots 10^{5}$ g~cm$^{-2}$, respectively,
where $\theta$ is the
incident angle of a cosmic ray into a cloud and 
$\langle\cos\theta\rangle=2/3$ for random injection.
\placefigure{fig:f1}
\placefigure{fig:f2}

\subsection{Simulations}
\begin{figure} 
\plotone{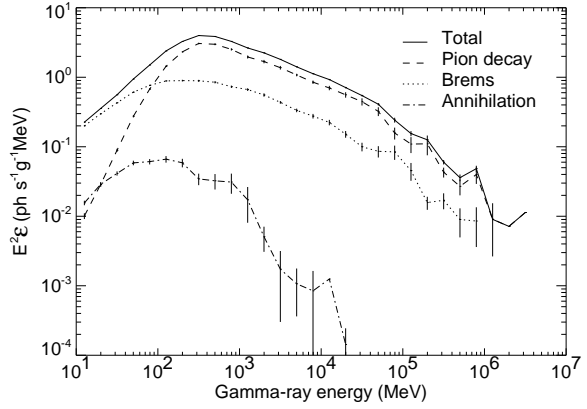}
\caption{Gamma-ray emissivity, $E^2{\cal E}$,
for a molecular hydrogen cloud,
of radius 1 AU and column density $100\;{\rm g\,cm^{-2}}$,
irradiated by cosmic-ray protons; vertical bars represent
Monte Carlo statistical errors. The various contributing
emission processes are also shown separately.
\label{fig:f3} } 
\end{figure}

Figure \ref{fig:f3} shows the resulting gamma-ray spectra
obtained in our GEANT4 simulation for proton injection into
a cloud of column density $\Sigma=100\;{\rm g\,cm^{-2}}$.
(The quantity plotted is $E^2{\cal E}$, with ${\cal E}$
being the emissivity as given in \S3.)
The dashed, dotted and dot-dashed lines
show the spectral components classified by the parent 
processes producing gamma-rays, i.e., $\pi^0$ decay,
bremsstrahlung and positron-electron annihilation, respectively.
The latter two components are necessarily omitted in
calculations which assume single interactions only
(i.e. the thin material limit).
The error bars are calculated from Monte Carlo statistics.
Although the $\pi^0$ decay component shows a broad
peak at $E\simeq70$~MeV (note that the quantity plotted
is the emissivity multiplied by $E^2$) and dominates above
about 200 MeV, the electron
bremsstrahlung component broadens the emissivity peak.
The bremsstrahlung and annihilation contributions
are negligible in the limit of small column-density;
their fractional contribution to the emission is
greatest for clouds which are just thick enough
($\Sigma\sim10^2-10^3\;{\rm g\,cm^{-2}}$) to
attenuate the bulk of the incident proton power.
This is as expected considering that the interaction
mean free path of GeV protons is $\sim80$~g~cm$^{-2}$
and is in accordance with the result of \citet{Ume81}
who treated a similar problem by solving one-dimensional 
transport equations.
 
\begin{figure} 
\plotone{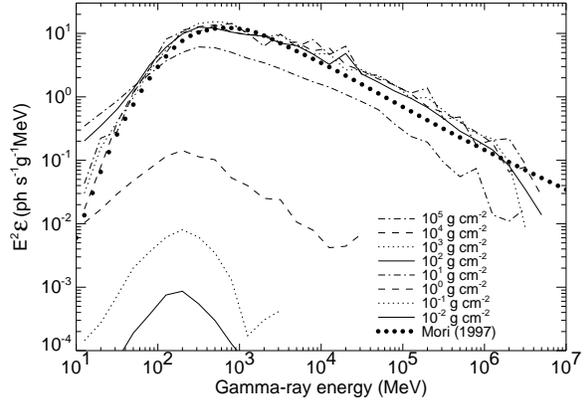}
\caption{Gamma-ray emissivities for dense clouds
irradiated  by cosmic-ray protons, with a spectrum appropriate
to the solar neighbourhood. The results have been multiplied
by a nuclear enhancement factor of 1.52 \citep{Mor97} to
account for the presence of heavier nuclei in the cosmic-rays.
The different curves are for various different column-densities:
$10^{-2}\,{\rm g\,cm^{-2}}$, thick dot-dashed curve;
$10^{-1}\,{\rm g\,cm^{-2}}$, thick dashed curve;
$1\,{\rm g\,cm^{-2}}$, thick dotted curve;
$10\,{\rm g\,cm^{-2}}$, thick solid curve;
$10^2\,{\rm g\,cm^{-2}}$, thin dot-dashed curve;
$10^3\,{\rm g\,cm^{-2}}$, thin dashed curve;
$10^4\,{\rm g\,cm^{-2}}$, thin dotted curve;
$10^5\,{\rm g\,cm^{-2}}$, thin solid curve. 
Also plotted is the emissivity from \citet{Mor97}, which corresponds
to the ``thin material'' limit (filled circles); this limit
offers a good approximation for $\Sigma\la10\,{\rm g\,cm^{-2}}$.
\label{fig:f4}}
\end{figure}

\begin{figure} 
\plotone{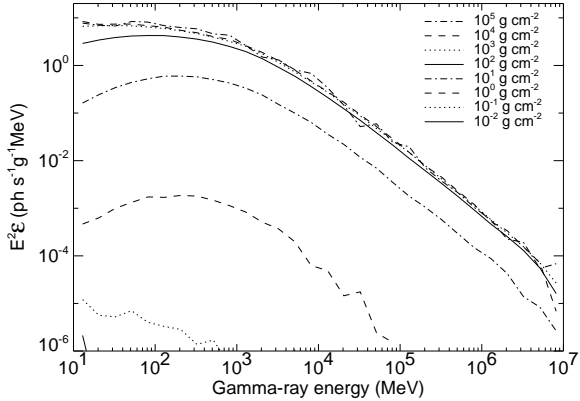}
\caption{Gamma-ray emissivities for dense
clouds irradiated  by cosmic-ray electrons. The different
curves are for different column-densities, as per figure
\ref{fig:f4}.
\label{fig:f5} }
\end{figure}

\placefigure{fig:f3}
\placefigure{fig:f4}
\placefigure{fig:f5}

The resulting gamma-ray emissivities for clouds of
various column densities are shown
in figure \ref{fig:f4} (proton injection)  and figure \ref{fig:f5}
(electron injection). 
Here the emissivities are defined for irradiation by
cosmic-rays of all species (see \S2.3, equation (3));
to take account of the contribution of heavier nuclei than helium,
the emissivity due to proton irradiation
(figure 4) has been multiplied by a nuclear enhancement
factor \citep{Cav71, Ste81, Der86, Gai92}
of 1.52 \citep{Mor97}. Note that for high densities the
Monte Carlo statistics are rather poor, since the yield itself is
low.  Figure \ref{fig:f4} includes a comparison of our
calculated gamma-ray production functions with that
of \citet{Mor97} (corresponding to the
``thin material'' limit). The results are consistent with
those of \citet{Mor97} for column densities less than about
$10$ g~cm$^{-2}$, except in the energy range
$E>10^6$ MeV where the effect of
the maximum energy assumed in the Monte Carlo simulation
is evident. We note the very low values of the emissivity at energies
$\ga100$~MeV, for column densities $\Sigma\ga10^3\,{\rm g\,cm^{-2}}$.
A slightly steeper spectrum in the $10^4$--$10^6$ MeV
region comes from our omission of the contribution of heavy nuclei,
which were taken into account in \citet{Mor97}. A somewhat
surprising feature of these curves is that the power-law
index above 1 GeV is almost the same as the input cosmic-ray
proton flux for column densities less than about 1000 g~cm$^{-2}$
(for higher column densities the statistics of the simulations
are not good enough to decide whether this result still holds).
This is already indicated by \citet{Ume81}, but is
contrary to the expectation of \citet{Sci00a} who
suggested a spectral change above 1 GeV at the point
where  self-shielding becomes important.

\begin{figure} 
\plotone{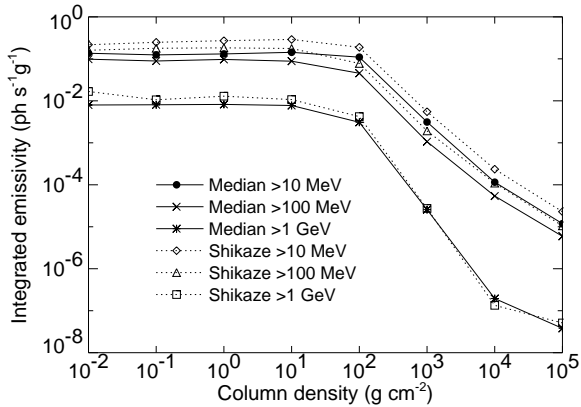}
\caption{Integrated gamma-ray emissivities for dense clouds
irradiated by cosmic-ray hadrons.
Solid curves are those assuming a cosmic ray proton flux
in \citet{Mor97} and dotted ones assuming that in \citet{Shi01}.
Here we multiplied by a nuclear enhancement factor of 1.52
\citep{Mor97} to take account of heavy nuclei in the cosmic rays.
\label{fig:f6} }
\end{figure}

\begin{figure} 
\plotone{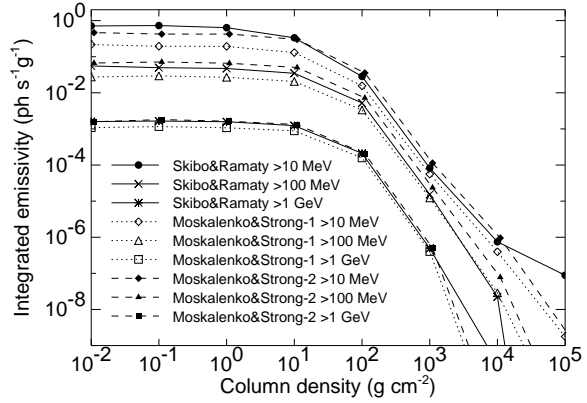}
\caption{Integrated gamma-ray emissivities for dense clouds
irradiated by cosmic-ray electrons.
Solid curves are those assuming a cosmic ray electron flux
in \citet{Ski93}, and dashed (dotted) ones assuming that 
in \citet{Mos98} with (without) re-acceleration.
\label{fig:f7} }
\end{figure}

\begin{figure} 
\plotone{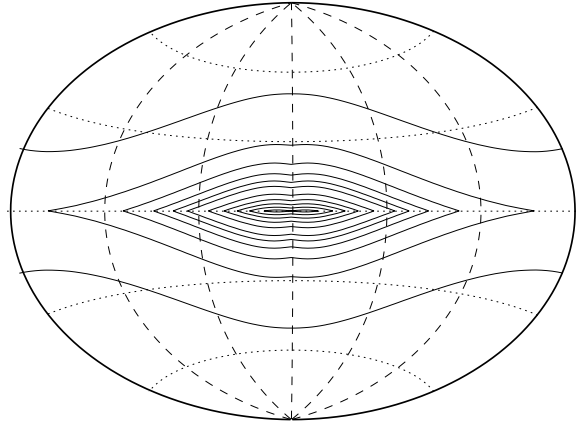}
\caption{Adopted model dark matter column-density distribution 
over the sky, $Q(l,b)$, weighted by cosmic-ray energy density and 
displayed in Galactic coordinates.  The quantity $Q$ is defined 
in equation 4 and is computed using the dark matter density model 
given in equation 5 together with the cosmic-ray distribution 
model given in equation 6.  Contour levels are 
$(4,8,12,16,20,24,32,40,48,56,64)\times10^{-3}\,{\rm g\,cm^{-2}}$; 
dashed lines are constant longitude, spaced at $60^\circ$ intervals; 
dotted lines are constant latitude, spaced at $30^\circ$ intervals .
\label{fig:f8} }
\end{figure}

For some purposes the energy-integrated emissivities
are of more interest than their differential counterparts, 
so we present these quantities in figures \ref{fig:f6}
and \ref{fig:f7}, for cosmic ray protons and electrons,
respectively.  At low column densities, where cascades
and self-shielding are unimportant,
there is very little variation of the integrated emissivity
with cloud column density, and the thin material limit
can be adopted for columns less than
$10\;{\rm g\,cm^{-2}}$ for protons ($1\;{\rm g\,cm^{-2}}$
for electrons).
Above this point, however, the emissivity drops
rapidly with increasing cloud column-density.
In order to gauge the sensitivity of these calculations to
the assumed cosmic-ray spectra, we have computed our
results for two different incident cosmic-ray proton
spectra, and three different incident cosmic-ray electron
spectra, as shown in figures 6 and 7, respectively.
The differences are
seen to be small in comparison with the variation as a
function of cloud column density, but at a fixed column-density
the systematic uncertainties associated with the input
cosmic-ray spectra are nevertheless significant.
\placefigure{fig:f6}
\placefigure{fig:f7}
\placefigure{fig:f8}

\begin{table}
 \centering
 \begin{minipage}{80mm}
  \caption{Integrated gamma-ray emissivities of dense gas clouds, due to cosmic-ray
  hadrons and electrons; these results assume Mori's (1997) ``median'' proton
  spectrum and Skibo and Ramaty's (1993) electron spectrum. A nuclear
  enhancement factor of 1.52 has been applied to the emissivities computed
  for cosmic-ray protons in order to account for heavier nuclei. Column-densities
  ($\Sigma$) are given in units of ${\rm g\,cm^{-2}}$ and integrated emissivities
  in units of ${\rm photons\,s^{-1}g^{-1}}$; photon energies ($E$) are given here in MeV.}
  \begin{tabular}{@{}llll@{}}
  \\
  \hline
  \\
   $\Sigma$ & $\int_{100}^\infty\!{\rm d}E\;{\cal E}$& $\int_{300}^{500}\!{\rm d}E\;{\cal E}$&$\int_{1000}^\infty\!{\rm d}E\;{\cal E}$ \\
   \\
 \hline
 $10^{-2}$& $1.47\times10^{-1}$ & $2.36\times10^{-2}$ & $8.22\times10^{-3}$  \\
 $10^{-1}$& $1.50\times10^{-1}$ & $2.34\times10^{-2}$ & $1.14\times10^{-2}$  \\
 $10^{0}$& $1.51\times10^{-1}$ & $2.61\times10^{-2}$ & $9.63\times10^{-3}$  \\
 $10^{1}$& $1.34\times10^{-1}$ & $2.15\times10^{-2}$ & $9.26\times10^{-3}$ \\
 $10^{2}$& $5.66\times10^{-2}$ & $9.04\times10^{-3}$ & $3.55\times10^{-3}$  \\
 $10^{3}$& $5.16\times10^{-3}$ & $8.43\times10^{-4}$ & $3.42\times10^{-4}$  \\
 $10^{4}$& $6.82\times10^{-5}$ & $7.11\times10^{-6}$ &  $4.81\times10^{-7}$ \\
 $10^{5}$& $6.65\times10^{-6}$ & $6.90\times10^{-7}$ &  $3.71\times10^{-8}$ \\
\hline
\end{tabular}
\end{minipage}
\end{table}

\begin{table}
 \centering
 \begin{minipage}{80mm}
  \caption{Gamma-ray emissivities of dense gas clouds, due to cosmic-ray
  hadrons and electrons; these results assume Mori's (1997) ``median'' proton
  spectrum, and Skibo and Ramaty's (1993) electron spectrum. A nuclear
  enhancement factor of 1.52 has been applied to the emissivities computed
  for cosmic-ray protons, in order to account for heavier nuclei. Column-densities
  ($\Sigma$) are given in units of ${\rm g\,cm^{-2}}$, and emissivities in units
  of ${\rm MeV\,s^{-1}g^{-1}}$; photon energies ($E$) are given here in MeV.
  The quantity ${\cal L}\equiv{\rm Log}_{10}(E^2{\cal E})$.}
  \begin{tabular}{@{}lllll@{}}
  \\
  \hline
  \\
   $E=$&10&100&1000&10000 \\
   \\
   $\Sigma$ & ${\cal L}$& ${\cal L}$&${\cal L}$&
   ${\cal L}$\\
 \hline
 $10^{-2}$&$0.889$&$1.086$&$1.126$&$0.657$\\
 $10^{-1}$&$0.924$&$1.059$&$1.198$&$0.763$\\
 $10^{0}$&$0.824$&$1.030$&$1.171$&$0.594$\\
 $10^{1}$&$0.497$&$0.961$&$1.145$&$0.598$\\
 $10^{2}$&$-0.223$&$0.584$&$0.745$&$0.164$\\
 $10^{3}$&$-1.356$&$-0.481$&$-0.276$&$-0.869$\\
 $10^{4}$&$-3.234$&$-2.175$&$-2.811$&$-4.628$\\
 $10^{5}$&$-4.251$&$-3.207$&$-3.928$&$-6.024$\\
\hline
\end{tabular}
\end{minipage}
\end{table}

\subsection{Computation of gamma-ray intensities}
Using the gamma-ray production spectra obtained
in the previous section, we have calculated the
diffuse gamma-ray emission from the Galaxy
as follows.
The predicted gamma-ray spectrum for each case is 
\begin{equation}
I_D = {1\over\Sigma}\int_0^\infty\!\!{\rm d}s\, \rho(s)
  J_{\rm cr}(s){{\rm d}N \over {\rm d}E}
\end{equation}
where ${\rm d}N/{\rm d}E$ is the spectrum returned
by the simulation in units of photons/MeV/primary,
for an individual cloud, appropriate to the incident
cosmic-ray spectrum. The quantity $J_{\rm cr}(s)$ is the intensity
of cosmic rays at a distance $s$ along the line of sight,
in units of primaries cm$^{-2}$~s$^{-1}$~sr$^{-1}$,
and $\rho(s)$ is the mean density in gas clouds of
column density $\Sigma$. 

The Galactic variation of the spectrum $J_{cr}(s)$ is
not well constrained by existing data, and consequently
we adopt the simplifying assumption that the shape of
the cosmic-ray spectra (both electrons and protons) is
the same everywhere in the Galaxy, with variations only
in the normalisation. With this assumption it is convenient
to recast the calculation as
\begin{equation}
I_D = {1\over{4\pi}} {\cal E} Q,
\end{equation}
(${\rm ph\,cm^{-2}s^{-1}sr^{-1}MeV^{-1}}$), where the
emissivity is
\begin{equation}
{\cal E} = {{4\pi}\over\Sigma}J_{\rm cr}(\odot) {{\rm d}N \over {\rm d}E}
\end{equation}
(${\rm ph\,s^{-1}MeV^{-1}g^{-1}}$)
with $J_{cr}(\odot)$ the cosmic-ray mean
intensity in the Solar neighbourhood and
\begin{equation}
Q \equiv \int_0^\infty\!\!{\rm d}s\, \rho(s)
  {{J_{\rm cr}(s)}\over{J_{\rm cr}(\odot)}}
\end{equation}
is the weighted column density (${\rm g\,cm^{-2}}$)
of the cloud population along the line-of-sight
under consideration. This formulation is convenient because
the emissivity, ${\cal E}$, describes the properties of the gas
clouds themselves and is independent of the Galactic
variations in mean dark matter density and cosmic-ray
density; conversely the quantity $Q$ characterises these
properties of the Galaxy, and is independent of the properties
of the gas clouds themselves. The emissivity shown in figures 4 and 5
is the quantity $E^2{\cal E}$, whereas figures 6 and 7 show
$\int_E^\infty\! {\rm d}E^\prime\,{\cal E}$. For the inner Galactic disk,
where we are interested in $\langle I_D\rangle$, we need to
average over the whole solid angle, $\Omega$, under
consideration: $\langle Q\rangle=\int\!{\rm d}\Omega\; Q\,/\Omega$.
In order to calculate $Q$ we need to adopt models for both
the Galactic cosmic-ray distribution and the Galactic
distribution of the clouds.

The quantity $\rho(s)$, the density in cold, dense
gas clouds,  is only weakly constrained by direct observation,
because the hypothetical clouds constitute a form of {\it dark\/} matter.
We therefore proceed by adopting a conventional dark matter
density distribution for the Galaxy, namely a
cored isothermal sphere, as our model cloud density distribution,
with a fiducial normalisation which is equivalent to the assumption
that all of the dark matter is in the form of dense gas clouds.
This corresponds to the model
\begin{equation}
\rho={{\sigma^2}\over{2\pi G(R^2+z^2+r_c^2)}},
\end{equation}
in terms of cylindrical coordinates $(R,z)$, with
$\sigma=155\;{\rm km\,s^{-1}}$. We have adopted a
core radius of $r_c=6.2$~kpc based on the preferred
model of Walker (1999). (This choice corresponds to
Walker's preferred value of cloud column density
$\Sigma=140\,{\rm g\,cm^{-2}}$.)  Walker's model
exhibits a core radius which is a function of cloud
column density, but we have fixed the core radius
at 6.2~kpc for all of our computations. This choice
permits more straightforward consideration of the
observational constraints because $Q$ is independent
of $\Sigma$ in this case.

It then remains to specify the cosmic-ray energy-density
as a function of position in the Galaxy. \citet{Web92}
(hereafter WLG92) constructed numerical models of
cosmic-ray propagation in the Galaxy; they did not give
any analytic forms for their model cosmic-ray distributions,
but an appropriate analytic approximation can be
deduced from the results which they obtained. They found
that the cosmic-ray radial distribution reflects,
in large part, the radial dependence of cosmic-ray
sources, with a modest smoothing effect introduced
by diffusion. We have therefore adopted WLG92's preferred
model (their model 3) for the radial distribution of
sources as our model for the radial distribution of
cosmic rays.

The various spectra of cosmic-ray isotope ratios
considered by WLG92 favour models in which the diffusion
boundaries are in the range $2-4$~kpc above and below
the plane of the Galaxy. We adopt the midpoint of this range.
WLG92 do not give a simple functional form for the vertical
variation of cosmic-ray density within this  zone, so we have
simply assumed an exponential model: $\exp(-|z|/h)$. We
know that in WLG92's models the cosmic-ray density is fixed
at zero at the diffusion boundaries, and consequently the
scale height of the exponential should be approximately half 
the distance to the diffusion boundary, i.e. $h\simeq1.5$~kpc.

These considerations lead us to the model cosmic-ray mean
intensity distribution $J_{cr}(R,z)$:
\begin{equation}
{{J_{cr}(R, z)}\over{J_{cr}(\odot)}}=\left({R\over{R_0}}\right)^{0.6}
 \exp[(R_0-R)/L -|z|/h],
\end{equation}
in terms of cylindrical coordinates $(R, z)$.  Here
$R_0\simeq8.5$~kpc is the radius of the solar circle,
while $L=7$~kpc, $h=1.5$~kpc and
$J_{cr}(\odot)=J_{cr}(R_0, 0)$. This distribution
has the character of a disk with a central hole.

These models for $\rho(s)$ and $J_{cr}$ allow us to compute
the quantity $Q$, as per equation 4, and the resulting variation over
the sky is plotted in figure 8. For reference we give the
values of $Q(l,b)$ evaluated at the cardinal points, as follows:
$Q(0, 0)=6.47\times10^{-2}{\rm g\,cm^{-2}}$,
$Q(\pm90^\circ,0)=1.54\times10^{-2}{\rm g\,cm^{-2}}$,
$Q(180^\circ,0)=7.70\times10^{-3}{\rm g\,cm^{-2}}$,
and $Q(b=\pm90^\circ)=2.44\times10^{-3}{\rm g\,cm^{-2}}$.
In order to compare with the EGRET results of \citet{Hun97},
we have also evaluated the average of $Q$ over the
inner Galactic disk:
$\langle Q(|l|\le60^\circ,|b|\le10^\circ)\rangle=
3.28\times10^{-2}{\rm g\,cm^{-2}}$.

\section{Discussion}

\subsection{Comparison with data}
Several physical processes contribute to the observed diffuse
gamma-ray intensity. In order of decreasing fractional
contribution these are thought to be pion production
and bremsstrahlung from cosmic-ray interactions with
{\it diffuse\/} Galactic gas, inverse Compton emission
from cosmic-ray electrons interacting with ambient
photons, and an isotropic background, which is presumably
extragalactic and due to many faint, discrete sources.
The sum of these contributions offers a good model for
the diffuse emission which is actually observed in the
100~MeV--1~GeV band \citep{Blo89, Hun97}; however,
the $E>\,1$~GeV emission is poorly modeled \citep{Hun97}
with the prediction \citep{Ber93} amounting to only $\sim2/3$ of
the observed intensity.

Some authors \citep{Gil94, Sal96} have used the low-energy
($E<\,1$~GeV) data to argue that the agreement between model
and data allows no room for any significant unmodeled emission
and that these data therefore place tight constraints on any
baryonic contribution to the dark matter halo of the Galaxy.
This line of argument is clearly suspect because of the
failure of the same model to account for the high-energy
($E>\,1$~GeV) data. However, even if we ignore the high-energy
data the calculations presented in \S2.3 demonstrate that
the gamma-ray constraints on high column density gas
clouds in the dark halo are much weaker than those on
low column density gas because the emissivity of the
latter is much greater; consequently, we revisit the published
constraints in the following section (\S3.2), returning to
the question of the high-energy data in \S3.3.

\subsection{Published constraints applied to high
column-density gas}
\citet{Gil94} argued that the known (diffuse) gas
accounts for essentially all of the gamma-ray
intensity in the $E>100$~MeV band observed by the
COS B satellite \citep{Blo89}, and he suggested
that any contribution from baryonic clouds should amount to
no more than $10^{-5}\,{\rm ph\,cm^{-2}s^{-1}sr^{-1}}$
($E>100$~MeV) at high Galactic latitudes.
Toward the Galactic poles we found (\S3)
$Q\simeq2.5\times10^{-3}{\rm g\,cm^{-2}}$, implying
that the average emissivity of the material in the dark halo
should be ${\cal E} < 5.0\times10^{-2}\,{\rm ph\,s^{-1}g^{-1}}$
for $E>100\;{\rm MeV}$.
From Table 1 (or figures 6 and 7) we see that this
requirement is not met for the models which are
in the thin material limit but that any model with
$\Sigma\ga100\,{\rm g\,cm^{-2}}$ is acceptable.
In other words the constraint imposed by \citet{Gil94}
permits the Galaxy's dark halo to be entirely baryonic,
provided the individual components have a
column-density $\Sigma\ga100\,{\rm g\,cm^{-2}}$.

\citet{Sal96} pointed out that obervations at low
Galactic latitudes can provide tighter constraints
on any baryonic dark halo, because of the higher
intensity expected when looking through the
cosmic-ray disk edge-on. This point is manifest
in the large values of $Q(b=0)$, which we computed
in \S2.3 (see also figure 8). \citet{Sal96} argued that a suitable
constraint on any baryonic component of the 
dark halo is that it should not contribute more than
the uniform background intensity component observed
in any given field. The strongest constraint then
comes from observations of the Ophiuchus region
\citep{Hun94}, for which the requirement imposed by
\citet{Sal96} is an intensity contribution from the
dark matter halo of
$I_D<2.4\times10^{-6}{\,\rm ph\,cm^{-2}s^{-1}sr^{-1}}$
in the band 300\ MeV $\le E \le$ 500\$ MeV.  For this
line of sight our calculation yields
$Q(0, b=15^\circ)=1.66\times10^{-2}{\rm g\,cm^{-2}}$,
and in the thin material limit, for which we find
${\cal E}\simeq2.4\times10^{-2}\,{\rm ph\,s^{-1}g^{-1}}$
(300\ MeV $\le E \le$ 500\$ MeV; see Table 1), this
corresponds to a predicted intensity of 
$I_D\simeq3.2\times10^{-5}{\,\rm ph\,cm^{-2}s^{-1}sr^{-1}}$
in the 300--500~MeV band, implying that at most 8\%
of the dark halo can be resident in low column density
clouds.\footnote{This result is a factor of 2 larger than the
corresponding model of \citet{Sal96} (with $L=3$~kpc
and a spherical halo), a difference which is accounted for
by their choice of a smaller core radius for the dark halo.
\citet{Sal96} chose a core radius of 3.5~kpc, which
yields $Q(0, b=15^\circ)=3.0\times10^{-2}{\rm g\,cm^{-2}}$,
whereas we have employed a core radius of 6.2~kpc
(see \S2.3).} However, our calculations  extend to
clouds of higher column-densities, and we find that
for $\Sigma\ga500\,{\rm g\,cm^{-2}}$ the limit relaxes
to the point where all of the dark halo is permitted to
be in the form of high column density clouds.

\subsection{Revision of constraints based on the observed spectrum}
\subsubsection{Galactic diffuse gamma-ray model}
The constraints discussed in the previous subsection make
reference only to the low-energy ($E<\,1$~GeV) gamma-ray
data. As noted earlier in this section, the observed intensity of
the inner Galactic disk at high photon energies ($>\,1$~GeV) is
substantially greater than expected \citep{Hun97, Ber93}.
Much effort has been expended on explaining this discrepancy,
with most of the attention given to models in which the Galactic
cosmic-ray electron and/or proton spectra differ from their
locally measured values \citep{Bus01, Aha00, Ber00, Str00, Mor97}.
However, none of these models offers a satisfactory explanation of
the observed mean gamma-ray spectrum of the Galactic disk,
and consequently a successful match to the low energy data
should not be taken to mean that the emission model is correct.
In turn this suggests that the constraints formulated by \citet{Gil94}
and \citet{Sal96}, on the basis of the low-energy data, may be
too restrictive. 

The question then arises as to what constraints the gamma-ray
data do in fact place on unmodeled emission, given the current
state of understanding of the observed emission.

The Galactic diffuse emission model used in \citet{Hun97}
is based on \citet{Ber93}, and contains the following contributions:
\begin{equation}
I = I_{\rm HI}+I_{\rm HII}+I_{\rm H_2}
  + I_{\rm IC}+I_{\rm EG},
\end{equation}
where $I_{\rm HI}$ is the gamma-ray intensity contributed by
cosmic-ray interactions with diffuse atomic hydrogen,
and similarly for the ionised and molecular components of
the interstellar medium (HII and ${\rm H_2}$, respectively).
The emissivity for these components \citep{Ber93} is, of course,
computed in the thin material limit, and each component therefore
has the same spectral shape, differing only in normalisation.
Here $I_{\rm IC}$ is the gamma-ray flux by inverse Compton emission
(cosmic-ray electrons up scattering low-energy photons),
and $I_{EG}$ is the isotropic extragalactic background flux.

Although the atomic and ionised hydrogen components can
be observed directly via their line emission, and are thus
well constrained, this is not true for the molecular
component. The molecular hydrogen column is assumed to be
proportional to the CO emission line strength, as measured
by the Columbia CO survey \citep{Dam87}, for example, but
the constant of proportionality (usually denoted by $X$) is
unknown and one is forced to determine its value by fitting to
the gamma-ray data.  Although the uncertainty in the best-fit
determination of $\langle X\rangle$ (averaged over the whole
sky) is small, the systematic uncertainties are acknowledged
to be much larger, ``at least 10\%--15\%''  \citep{Hun97}.  In turn,
this estimate of the uncertainty is small in comparison with the
differences among the various values of $X$ which have
been reported in the literature (see the review by \citet{Blo89})
and the likely range of variation in $X$ within the Galaxy
\citep{Hun97}. Although these are substantial uncertainties,
molecular hydrogen contributes only a fraction of the total
observed gamma-ray intensity roughly 20\% of the
local surface density of gas in the Galactic disk is thought
to be in molecular form \citep{Wou90} so
the implied fractional uncertainty in the total
Galactic emission is perhaps as small as 3\%. Moreover,
uncertainty in $X$ affects only the normalisation and not
the spectral shape of the predicted emission from diffuse
molecular hydrogen, so the observed high-energy excess
cannot be explained in this way even if $X$ could assume
an arbitrarily large value.

The other main contributions to uncertainty in the diffuse model
are associated with (1) unmodeled spatial variations in the cosmic-ray
spectral energy density and (2) unmodeled spatial variations in the
low-energy photon spectral energy density; the former affects all of
the Galactic contributions to $I$, whereas the latter affects only the
inverse Compton component. These uncertainties affect both
the normalisation and the spectral shape of the predicted gamma-ray
emission; however, the uncertainties are difficult to quantify.

\subsubsection{Constraints based on the gamma-ray data}
For our purposes it is not actually necessary to quantify the
uncertainties on the model input parameters; it suffices
to use the discrepancy between model and data as
a measure of the uncertainty in our understanding of the
observed emission. In turn this measure determines the
constraints which we can apply to any putative unmodeled
emission, such as the contribution from dense gas which
we are concerned with here. At photon energies $E>1\;{\rm GeV}$
the fractional discrepancy is roughly 60\% \citep{Hun97},
in the sense that the observed emission is 1.6 times larger
than the model, and we henceforth adopt $0.6/1.6\simeq40$\%
of the total observed intensity as our estimate of the unmodeled
emission. Although this estimate is derived from data at high energies,
the effects of the various contributing processes are 
all very widely spread, and
{\em the estimate therefore applies independent of photon energy.\/} 
The constraints appropriate to high/low
Galactic latitudes can now be re-evaluated.

At high Galactic latitudes the observed intensity is
$I\simeq1.5\times10^{-5}\,{\rm ph\,cm^{-2}s^{-1}sr^{-1}}$
for $E\ge100\;{\rm MeV}$
\citep{Kni96}, implying that any unmodeled emission should be
$\la6\times10^{-6}\,{\rm ph\,cm^{-2}s^{-1}sr^{-1}}$ in this band.
This result is actually slightly stricter than the criterion used by
\citet{Gil94} and thus leads us to tighten our high-latitude
constraints, relative to those quoted in \S3.2: the observed
high-latitude gamma-ray intensity constrains the amount of
low column-density gas to $\la20$\% of the total density of the
Galactic dark halo, with this fraction rising to 100\% for gas
clouds of column density $\Sigma\ga200\;{\rm g\,cm^{-2}}$.

At low Galactic latitudes we can make use of the mean
intensity of the inner Galactic disk, which has been accurately
determined by \citet{Hun97}. For example at 1~GeV the
mean intensity ($|l|\le60^\circ$, $|b|\le10^\circ$) is
$\langle I\rangle\simeq
3\times10^{-8}\,{\rm ph\,cm^{-2}s^{-1}sr^{-1}MeV^{-1}}$,
and our calculation of $\langle Q\rangle$ for this region yields
(\S2.3) $3.28\times10^{-2}\,{\rm g\,cm^{-2}}$, implying that the
emissivity of the Galactic dark halo material must be, on average,
${\cal E}\le4.6\times10^{-6}\,{\rm ph\,s^{-1}g^{-1}MeV^{-1}}$.
By comparison, the actual emissivity of low column-density
gas is computed to be (\S2.3, table 2) 
${\cal E}(1\;{\rm GeV})\simeq1.4\times10^{-5}\,{\rm ph\,s^{-1}g^{-1}MeV^{-1}}$,
implying that $\la30$\% of the Galaxy's dark halo may be
comprised of low column density gas. For higher column densities
the emissivity falls, and table 2 shows that for $\Sigma=100\,{\rm g\,cm^{-2}}$
the emissivity is only $5.5\times10^{-6}\,{\rm ph\,s^{-1}g^{-1}MeV^{-1}}$.
The gamma-ray data on the inner Galactic disk thus indicate all of the
Galaxy's dark halo to be made of dense clouds of column-density
$\Sigma\ga100\,{\rm g\,cm^{-2}}$.

\subsection{Comment on the gas content of the galactic disk}

The constraint we have just given is based on the mean spectrum
of the inner Galactic disk, in contrast to those given by \citet{Sal96}
who employed limits based on the angular structure of the observed
gamma-ray intensity. Specifically, \citet{Sal96} required that the
putative contribution of emission from a baryonic component of the
Galaxy's dark halo be less than that of the isotropic component of the
observed intensity; this procedure seems to us to be less reliable
than the procedure we have employed, for two reasons. First, even
if the dark halo were spherically symmetric the emission from any
baryonic component would not be, both because our point
of observation is quite
distant from the centre of the Galaxy and because the resulting
gamma-ray emission is strongly dependent on the Galactic cosmic-ray
distribution, which in turn is strongly concentrated in the disk of the Galaxy.
Indeed the cosmic-ray distribution appears to correlate with the
distribution of interstellar matter \citep{Hun97}, thus complicating the
interpretation of the observed gamma-ray intensity distribution. In
particular this coupling leads to gamma-ray emission from a
baryonic dark halo being correlated with the diffuse gas column density,
even if the dense gas is uncorrelated with the diffuse gas. 
Second, on any given line of sight, such as the Ophiuchus field
considered by \citet{Sal96}, a highly structured dark matter halo might
exhibit, by chance, a low dark matter column density. The chances
of this are good if, as in the case of \citet{Sal96}, the field is
specifically chosen to have a low ``background'' intensity.

\section{Summary}
The gamma-ray spectra arising from cosmic-ray interactions with
gas clouds of various column-densities have been calculated using
a Monte Carlo event simulator, GEANT4. Our calculations reproduce
the analytic result in the low column-density limit, where only
single particle interactions need to be considered, but exhibit
significant differences for clouds of column-density
$\Sigma\ga10^2\;{\rm g\,cm^{-2}}$ where the emissivity
declines substantially for photon energies $E\ga100\;{\rm MeV}$.
The low emissivity of dense gas means that the baryonic content
of the Galaxy's dark halo is not so tightly constrained by the gamma-ray
data as had previously been thought.  For $\Sigma\ga200\;{\rm g\,cm^{-2}}$
we find that the existing gamma-ray data, taken in isolation, do not
exclude purely baryonic models for the Galactic dark halo.

We are grateful to the referee for constructive comments that have
improved this paper.

\clearpage

\end{document}